\documentclass{iau_FM}
\usepackage{graphicx}

\title[OJ 287] %% give here short title %%
{The New Variability Phase of OJ 287 and Emergence of New Components in
NIR to X-ray Region} %% give here the full title %%

\author[Pankaj Kushwaha et al.]   %% give here short author list %%
{P. Kushwaha$^1$, E. M. de Gouveia Dal Pino$^1$, A. C. Gupta$^2$ \and P. J. Wiita$^3$
}

\affiliation{$^1$Department of Astronomy (IAG-USP), University of Sao Paulo, Sao Paulo 05508-090, Brazil
\\ email: {\tt pankaj.kushwaha@iag.usp.br} \\
$^2$Aryabhatta Research Institute of Observational Sciences (ARIES),
Manora Peak, Nainital 263002, India\\
$^3$Department of Physics, The College of New Jersey, P.O. Box 7718, Ewing, NJ 08628-0718, USA}

\pubyear{2018}
\jname{Astronomy in Focus, Volume 1} 
\editors{Volker Beckmann \& Claudio Ricci, eds.}
\begin{document}
\maketitle

\begin{abstract}
We present a multi-wavelength (MW) spectral and temporal study of the recent activity
of the claimed super-massive binary black hole system OJ 287 since December 2015.
The overall MW activity can be divided into two durations: December 2015 -- April
2016 (MJD: 57360 -- 57500), showing strong activity from near-infrared (NIR) to
$\gamma$-rays and September 2016 -- June 2017 (MJD: 57650 -- 57930), showing intense
NIR to X-ray variability concurrent with detection at very high energies (VHE) by
VERITAS, but without any signatures of variability in the Fermi-LAT band. In the
first duration, the variations are almost simultaneous and the SEDs show new components
in NIR-optical and optical-UV region. The NIR-optical bump is consistent with standard
accretion-disk (AD) description while the optical-UV appears consistent with
contributions from the broad-line region. The extracted broadband SEDs also show
a clear shift in gamma-ray SED peak and can be explained with inverse Compton
scattering of photons from broad line region. In the second period, the variations
are also simultaneous except for one duration during which X-ray leads the
optical/UV by $\sim$ 5-6 day. The broadband SEDs, on the other hand, show mixture of
a typical OJ 287 SED and an HBL SED, consistent with an origin from two different
zones, one located at sub-parsec scales and other at parsec scales.

%% add here a maximum of 10 keywords, to be taken form the file <Keywords.tex>
\end{abstract}

\firstsection % if your document starts with a section,
\section{Introduction}

Blazar are radio-loud active galactic nuclei (AGNs) with their jet align at a
close angle to the observer's line of sight. They are characterized by
a rapid variability with emission spanning the entire accessible electromagnetic
(EM) spectrum, high and variable polarization, and frequent detection of superluminal
motion. Temporally, the variability is consistent with stochastic variation while
the in the spectral domain, they exhibit a characteristic broad double hump spectral
energy distributions (SEDs). 
The lower energy hump is well understood be synchrotron emission from relativistic
non-thermal electron population while high energy hump can arise from inverse
Compton (IC) scattering of photons within the jet and/or surrounding
photons depending on the location of emission region e.g. broad line region and
torus IR photons at sub-parsec scales, torus IR photons at parsec
scales etc (e.g. see \cite{2013MNRAS.433.2380K}). In this scenario, the emission
at different wavelengths is highly correlated, both in the spectral and temporal
domain, due to being generated from the same underlying particle population.
Thus, any change in lower energy hump of SED should be reflected at the higher
energy hump as well.

\section{OJ 287 \& Recent MW Activity}
OJ 287 is an optically bright BL Lac object with optical record available since
1890. It shows regular quasi-periodic outbursts of $\sim$ 12 years in the optical
band in addition to the typical stochastic variability of blazars. The regular
feature has been suggested to be a result of precessing, binary supermassive black
hole (SMBH) system (\cite{2016ApJ...819L..37V}, and references therein). The model,
with many improvements since its inception, has been relatively successful in
predicting the timing of these regular outbursts and predicted the latest of
next outburst around December 2015 -- January 2016.

As per prediction, an increase in NIR-optical emission was observed at
November 2015 end. The predicted impact outburst was observed on December 5,
2016 (MJD 57361) with a relatively low optical polarization ($<$ 10\%,
\cite{2016ApJ...819L..37V}). However, a systematic rotation of $\sim$ 200$^\circ$
in optical polarization was observed (\cite{2017MNRAS.465.4423G}), untypical for
a dominant thermal emission as argued in the above model. Further, outburst at
X-rays and $\gamma$-rays was also seen. Following this till mid-2017, the source
has been very active across the entire EM spectrum which is the focus of our report here.

% \section{Multi-wavelength Activity, Results and Conclusion}
Temporally and spectrally, the overall MW activity can be clubbed into
two durations: December 2015 -- April 2016 (MJD $\sim$ 57300 -- 57500 ) and
June 2016 -- September 2017 (MJD $\sim$ 57650 -- 57930), as
is originally presented in \cite{2018MNRAS.473.1145K} and \cite{2018MNRAS.479.1672K}.
The first duration showed activity from NIR to Fermi-LAT $\gamma$-ray energies
while the second duration showed a historic NIR to X-ray activity but lacks 
variability at Fermi-LAT $\gamma$-ray energies. The historic phase coincided
with the first ever detection of OJ 287 at VHE energies by VERITAS
(\cite{2017arXiv170802160O}).

Correlation study shows that the variations are simultaneous except for one
period during the second duration which shows a systematic variation with optical/UV
lagging X-ray by $\sim$ 5-6 days. In the spectral domain, the first duration revealed
two new features: a bump at NIR-optical interface and a hardened Fermi-LAT $\gamma$-ray spectrum
with a shift in its peak location (e.g. \cite{2013MNRAS.433.2380K} for comparison).
The NIR-optical bump is consistent the standard Shakura-Sunyaev accretion disk description of
the primary SMBH ($1.8\times10^{10} M_\odot$) while the $\gamma$-ray spectral
hardening and peak shift can be naturally produced if it originates as a result of IC
scattering of BLR photons as reported in \cite{2010A&A...516A..60N}
for previous impact periods. This explanation resonates with the fact that
NIR-optical SED shows no sign of a shift in its peak location. The IC/BLR constraining
the emission region location at sub-parsec scales. The SEDs during the second
duration, on the other hand, show an additional non-thermal component peaking
at X-ray energies during high activity states while the X-ray spectra show curved
spectrum with departure consistent with the level of X-ray emission during the 
quiescent state of the first duration.
The location of X-ray peak and coincident VHE detection is consistent with properties
of HBL blazars and can be reproduced by an addition emission region at parsec
scales if emission is related with peak escape events of the binary SMBH model.

PK acknowledges FAPESP support grant 2015/13933-0.


\begin{thebibliography}{}

\bibitem[Gupta et al.(2017)]{2017MNRAS.465.4423G} {Gupta, A.~C., Agarwal, A., Mishra,
A., et al.}\ 2017, \textit{MNRAS}, 465, 4423

\bibitem[Kushwaha et al.(2018b)]{2018MNRAS.479.1672K} {Kushwaha, P., Gupta, A.~C.,
Wiita, P.~J., et al.}\ 2018b, \textit{MNRAS}, 479, 1672 

\bibitem[Kushwaha et al.(2018a)]{2018MNRAS.473.1145K} {Kushwaha, P., Gupta, A.~C.,
Wiita, P.~J., et al.}\ 2018a, \textit{MNRAS}, 473, 1145 

\bibitem[Kushwaha et al.(2013)]{2013MNRAS.433.2380K} {Kushwaha, P., Sahayanathan,
S., \& Singh, K.~P.}\ 2013, \textit{MNRAS}, 433, 2380

\bibitem[O'Brien et al. (2017)]{2017arXiv170802160O} {O'Brien S. for VERITAS
Colloboration} 2017, \textit{arXiv}, arXiv:1708.02160

\bibitem[Nilsson et al.(2010)]{2010A&A...516A..60N} {Nilsson, K., Takalo, L.~O.,
Lehto, H.~J., \& Sillanp{\"a}{\"a}, A.}\ 2010, \textit{A\&A}, 516, A60 

\bibitem[Valtonen et al.(2016)]{2016ApJ...819L..37V} {Valtonen, M.~J., Zola, S.,
Ciprini, S., et al.}\ 2016, \textit{ApJL}, 819, L37

\end{thebibliography}
\end{document}